\documentclass[aps,physrev,twocolumn,groupedaddress, showkeys, floatfix, nofootinbib]{revtex4-2}
\usepackage[pdftex]{graphicx}
\usepackage{amsmath,amsfonts}
\usepackage{csquotes}

\begin{document}

\preprint{}

\title{Cooling rate and glassy behavior in the Fermi--Pasta--Ulam system}

\author{M. Razza}
\affiliation{School of Physics, University of Milan, Via Celoria 16, 20133 Milano (Italy).}
\author{A. Carati}
\email{Corresponding author: andrea.carati@unimi.it}
\author{L. Galgani} 
\affiliation{Mathematics Department, University of Milan, Via Saldini 50, 20133 Milano (Italy).}

\date{\today}

\begin{abstract}
In this work, we numerically studied the cooling process of a
Fermi--Pasta--Ulam system, which occurs when the FPU system is placed
in contact with a gas whose temperature $T$ is reduced with a certain
cooling rate $\xi$. It was found that the existence of a weak
stochastic threshold has a significant impact on the cooling process,
because below such a stochastic threshold the specific FPU energy
is larger than its temperature, i.e., the FPU system falls out of
equilibrium. The difference remains finite in the limit $T \to 0$, so
that the FPU system maintains a residual amount of energy $E_0$ at
vanishing temperature. Our numerical simulations reveal that this
energy exhibits a power--law dependence on both the system size and
the cooling rate, scaling approximately as $E_{0} \sim (\xi N)^{2/3}$.
\end{abstract}


\maketitle

\section{\label{sec1}Introduction}
There exists a wide class of liquids that, if cooled sufficiently
fast, remain liquids also below the freezing point, becoming
super cooled liquids. If the cooling process is continued, one ends up
with a glassy solid. It is believed that (see for example \cite{EAN}
and the references within) this is due to the fact that the
thermalization time increases so rapidly as the temperature decreases
that any feasible cooling process becomes a non--equilibrium one.

Due to the great complexity of realistic models, little is known, at
an analytical level, about the dynamics of such systems. However, the
idea has been put forward very recently (see, for example,
\cite{DYZZX}) that this lack of a thermalization could be understood
in the frame of the theory of perturbations of Hamiltonian systems in
the thermodynamic limit, a theory that has been developed in recent
years \cite{C,MC,CM,GPP,M} and applied, for example, to the famous
Fermi--Pasta--Ulam model.

For this system, perhaps the simplest model of a solid, some
analytical results are known: for example, in the work \cite{BCM}, it
has been shown that the thermalization time diverges in the limit in
which the specific energy vanishes, and numerical estimates
\cite{BCP,Fu1,Fu2} show that such times grow at least as $u^{-2}$,
where $u$ is the specific energy.

Then it is clear that this system can be considered a good candidate
for the study of cooling processes involving glassy systems. So,
following the paper \cite{CG}, we numerically investigate the cooling
process obtained by placing the Fermi--Pasta--Ulam system in contact
with an ideal gas, the energy of which is gradually reduced.

This process has been studied by varying both the number of particles
$N$ of the system and the cooling rate $\xi$. The main results seem to
be two. First of all, there seems to be a temperature below which the
thermodynamic quantities suddenly cease to be those of equilibrium:
for example, the thermodynamic energy no longer coincides with the
mechanical energy and, even at extremely low temperatures, the
Fermi--Pasta--Ulam system maintains a quota of energy that we have
called the zero--point energy $E_0$. This phenomenon could be
analogous to the phenomenon for which the latent heat of
crystallization, in super cooled liquids, is not released to the
thermostat. In the glassy state, the solid thus have, at the same
temperature, having an energy higher than the energy of the
crystalline state.

The second result concerns the dependence of the zero--point energy
$E_0$ on the number of particles $N$ in the system and on the cooling
rate $\xi$: from the numerical simulations, it seems possible to
conclude that the first result cited may also remain at the
thermodynamic limit. This could be seen as a confirmation of the ideas
expressed in the already cited paper \cite{CG}.

This paper is organized as follows. In Section \ref{sec2} we describe
the model and the method we use to implement the cooling process,
while in Section \ref{sec3}, the results of the numerical simulations
are presented. Finally, the conclusions follow.

\section{\label{sec2}The model}
As we recalled in the Introduction, our aim is to study the cooling
process of a FPU system. This is a simple model of a crystal and
consists of a linear lattice of $N+2$ points, the first and last kept
fixed, while the other interact with first neighborhood
interactions. The Hamiltonian $H_{\mathrm{FPU}}$ of the system is
therefore written as
\begin{equation}\label{eq:1}
    H_{\mathrm{FPU}} = \sum_{j=1}^N \frac {p_j^2}{2m} + \sum_{j=0}^N
    V(q_{j+i}-q_j),
\end{equation}
\begin{figure}[htbp]
    \includegraphics[width=\columnwidth]{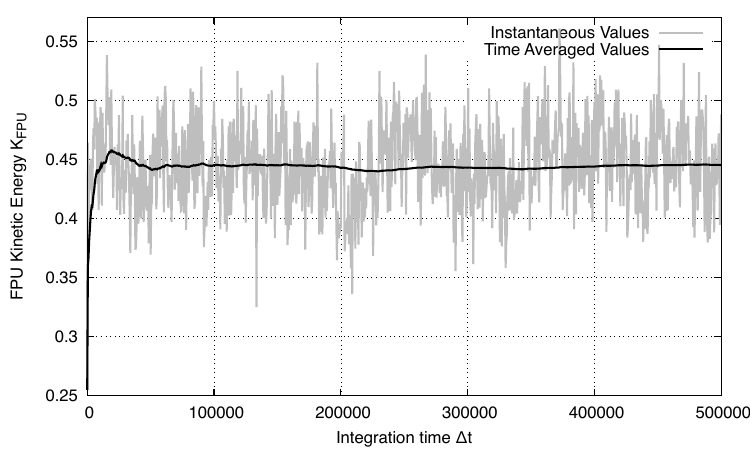}
    \\ \includegraphics[width=\columnwidth]{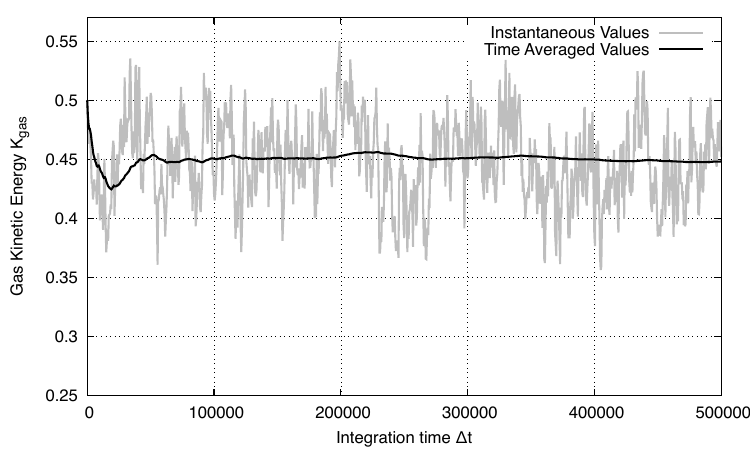}
    \caption{\label{keconv}FPU kinetic energy vs. time (upper panel)
      and gas kinetic energy per particle vs. time (lower one). Notice
      that the two kinetic energies essentially coincide after a time
      of order $10^5$.}
\end{figure}
whit $q_0=-L$, $q_{N+1}=0$, where $q_j$, $p_j$, $j=0 ,\ldots, N+1$ are
the position of the $j$--th particle and its conjugate momentum
respectively, while the parameter $L$, which determines the
equilibrium position of the particles\footnote{It is immediate to
verify that the equilibrium position corresponds to the particles
being $L/(N+1)$ apart.}, will be specified in a moment. Following the
reference \cite{BSBL}, the inter particle potential $V(r)$ is the
Lennard--Jones potential, i.e., we choose
\begin{equation}\label{eq:2}
  V(r)=4V_0\bigg( \Big(\frac {\sigma}r \Big)^{12} - \Big(\frac
  {\sigma}r \Big)^6\bigg) \ .
\end{equation}
The characteristic length $\sigma$ and the energy $V_0$ have the
following meaning: the potential vanishes for $r=\sigma$, while its
minimum is located at $\sigma^* := \sqrt[6]{2} \sigma$; instead, the
characteristic energy $V_0$ is the depth of the potential well.

Now, we come back to the parameter $L$, i.e. the length of the
Fermi--Pasta--Ulam system, that we choose as $L=(N+1)\sigma^*$, so
that at the equilibrium, the particles sit at the bottom of the
potential well, and the Hamiltonian gets the minimum possible value
$H_{\mathrm{FPU}}=-NV_0$. Notice that, in the following, the FPU
energy will be computed \textbf{with respect to this minimum}, i.e.,
adding the constant $NV_0$ to the expression (\ref{eq:1}), so it will
obviously be a positive quantity.

According to \cite{BSBL}, for a specific energy $u=E/N$ below
$V_0/27$, where the Fermi--Pasta--Ulam energy $E$ is calculated
according to the convention mentioned above, the Fermi--Pasta--Ulam
system shows an ordered behavior on a very long time scale. So, we
expect that some deviation from the equilibrium behavior could be
detected during the cooling process, once this threshold is reached.

As explained in the Introduction, the cooling process is performed by
putting in contact the Fermi--Pasta--Ulam system with a
one--dimensional perfect gas, with the same number of particles as the
Fermi--Pasta--Ulam system. We choose as a model of gas, a system of
$N+2$ non interacting particles\footnote{In any case, for collision
between two identical particles on a line, one gets simply an exchange
of the particles momenta, so that the interactions between particles
is not very effective, requiring at least ternary collisions.} of mass
$m$, the same as the particle mass of the FPU system, moving on the
positive half--line and confined in an interval of length $R$ by a
potential barrier, which interacts with the $N$--th particle of the
Fermi--Pasta--Ulam system through an interaction potential
$V_{\mathrm{int}}$. Therefore, the total system, Fermi--Pasta--Ulam
and gas, is Hamiltonian, the total Hamiltonian $H_{\mathrm{tot}}$
being
\begin{equation}\label{eq:3}
 H_{\mathrm{tot}} = H_{\mathrm{FPU}}(p_1,\ldots,q_{N+1}) +
 \sum_{j=0}^{N+1} \frac {\pi_j^2}{2m} + \sum_{j=0}^{N+1}
 V_{\mathrm{int}}(y_j-q_N) \ ,
\end{equation}
where $y_j$ and $\pi_j$, $j=0,\ldots,N+1$, are the position and
momentum, respectively, of the $j$--th gas particle, and it is
understood\footnote{So we do not write explicitly the gas confining
potential in the Hamiltonian.} that the gas particles suffer an
elastic collision at $y_j=R$. We take $R = 10L$, because, under
ordinary conditions, the distance between gas particles is an order of
magnitude greater than the distance between particles in a
solid. Concerning the interaction potential, we choose the following
form
\begin{equation}\label{eq:4}
  V_{\mathrm{int}}(r) = V_1 \frac{e^{-\alpha(r/\sigma)^2}}{r/\sigma}
  \ ,
\end{equation}
with parameters $V_1=100 V_0$ and $\alpha=50$. These values are not
realistic, but they have been chosen by trial and error to ensure that
we reach thermalization at high temperatures in a reasonable time.

We choose as units of length, mass and energy the quantity $\sigma$,
$m$, $V_0$ which are therefore set equal to 1 during the computations,
and moreover, we set the Boltzmann constant $k_B=1$, so that
temperatures are expressed in units of energy.

The equations of motions are integrated by the standard Verlet method,
which is symplectic and ensures a good conservation of the energy for
very long integration times too. The integration step $\tau$ is taken
equal to $0.0025$ in our time unit (which is given by
$\sqrt{m\sigma^2/V_0}$), to ensure a relative energy conservation
better than $0.01\%$.

The cooling process is performed as follows. To start with, we choose
initial data so that the gas particles are uniformly distributed in
the segment $[0, R]$ while their velocities are extracted from a
Maxwell distribution at a temperature $T=1$ (in our unit),
corresponding to a kinetic energy per particle half of the potential
well. For what concerns the FPU system, one should give the initial
data by taking them at random from a Gibbs distribution. We prefer,
instead, to start from a non--equilibrium distribution and let the
system equilibrate itself. In such a way, we can estimate the
thermalization time for our system, at least at high temperatures. So,
the FPU initial data are given by extracting the energy of the normal
modes by a Maxwell distribution at a temperature $T/2$.\footnote{We
choose $T/2$, and not $T$, because for $T=1$ the system is strongly
non--linear so that, after a small transient, the kinetic energy of
the FPU increases to a value close to $T$.}

We integrate until the two systems equilibrate. As can be seen in
Figure \ref{keconv}, which refers to an FPU system of $N+2=200$
particles and a temperature $T=1$, the kinetic energies of the two
systems (gas and FPU) averaged over a time interval $\Delta t= 5 \cdot
10^5$ (in our unit) are essentially equal after a time integration
less than $10^5$. As will appear more clear in the next Section, the
specific energy of the FPU system is lower than $T$, the value
expected for a linear system, which means that the non linear
contributions are relevant at this temperature.

After equilibration, we start the cooling process. This process is
simulated by a sequence of discrete steps. In each step, we decrease
the velocity of each gas particle, while the positions are left
untouched, and then we let the two systems equilibrate again for a
time $\Delta t$. We repeat this step a certain number of times
$k_{max}$ (which has been fixed equal to $90$) until we reach the
final temperature. The reduction in velocity is implemented by
computing a scale factor $\eta<1$ and simply multiplying the velocity
of the particles by this factor. The scale factor is computed as
\begin{equation}
  \eta = \sqrt{\frac {T_{new}}{T_{old}} },
\end{equation}
where $T_{old}$ is the actual temperature of the gas (see below) and
$T_{new}$ is the target temperature. The target temperature $T_{new}$
is obtained by decreasing $T_{old}$ of a fixed amount ($\Delta T=0.05$
in our case), for $T_{old}$ above a certain threshold, while it is
simply $3/4$ of $T_{old}$ below such threshold. In this way, we
simulate a cooling process with a fixed cooling rate up to a certain
temperature threshold, below which the rate decreases exponentially.
\begin{figure}[htbp]
    \includegraphics[width=\columnwidth]{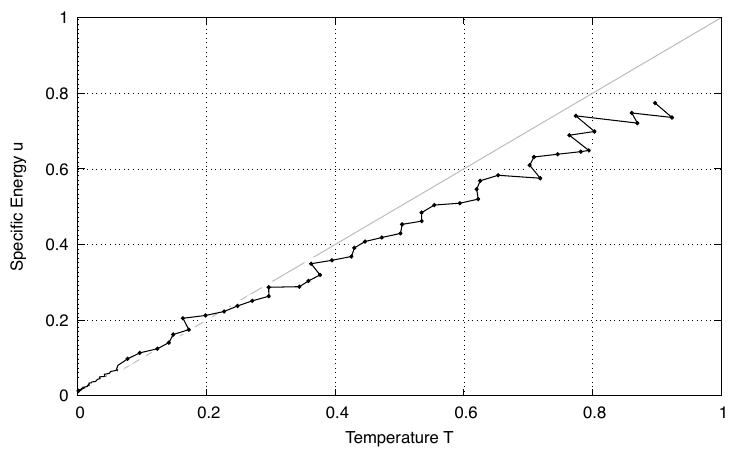}
    \\ \includegraphics[width=\columnwidth]{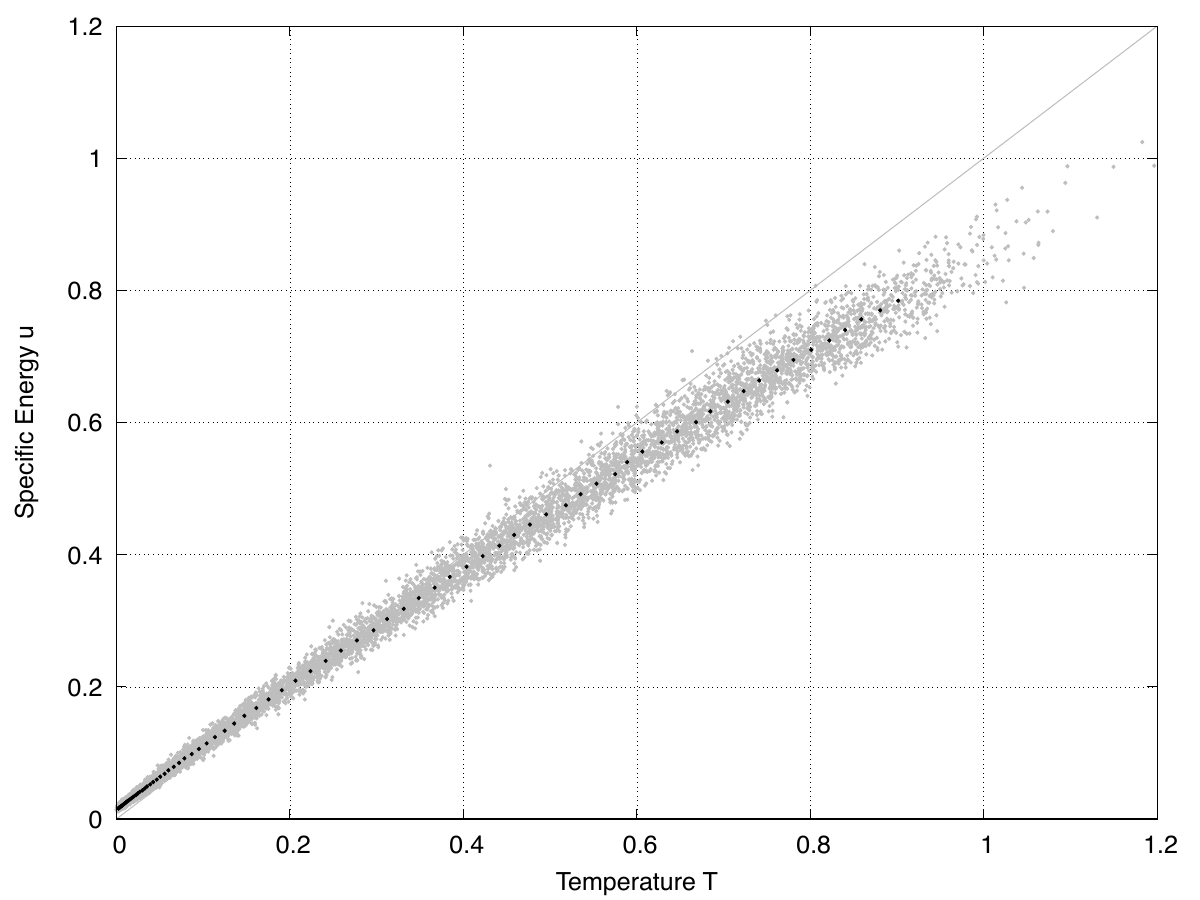}
    \caption{\label{cooling_process} Cooling process starting from
      $T_{gas} = 1$ with time step $\tau = 0.0025$, $N+2 = 200$,
      initial energy of normal modes $E_k \approx 0.5051$ for all
      k. Upper panel: single trajectory. Lower panel: 128
      trajectories. Grey points are data for the single trajectory,
      black ones are the resulting average.  }
\end{figure}

\section{\label{sec3}Numerical results}
We now come to the numerical results. The thermodynamic quantities of
interest, temperature $T$ and internal energy of the FPU system $U$,
are defined the first as the double of the average specific kinetic
energy of the gas, i.e. $T := \big\langle \sum
\pi_j^2/m\big\rangle/N$, while the second is defined as the average of
the FPU Hamiltonian, i.e. $U := \big\langle
H_{\mathrm{FPU}}\big\rangle$. The problem is now how to perform the
averages.

In the literature, the averages are often taken as the time averages
along an orbit followed by a further average over a certain number of
different orbits. The initial data for such orbits are so chosen as to
provide a good sample for the assumed distribution, canonical if
working at fixed temperature, or micro canonical if working at fixed
energy.

\begin{figure}[htbp]
    \includegraphics[width=\columnwidth]{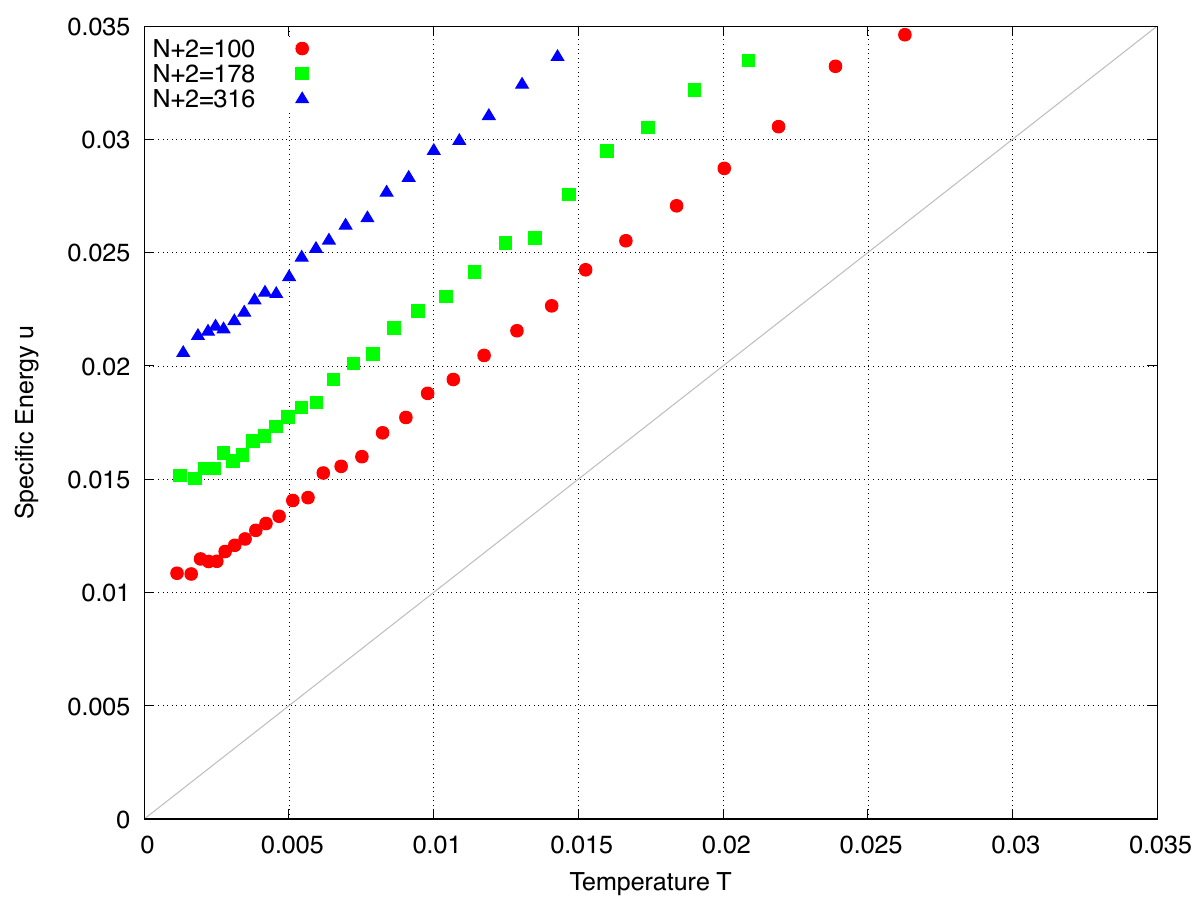}
    \caption{\label{fig:4} Cooling process for different values of
      $N+2 = 100,178,316$, for temperature below $0.035$. Here the
      cooling rate is $\xi=4 \cdot 10^{-7}$.  }
\end{figure}
\begin{figure}[htbp]
    \includegraphics[width=\columnwidth]{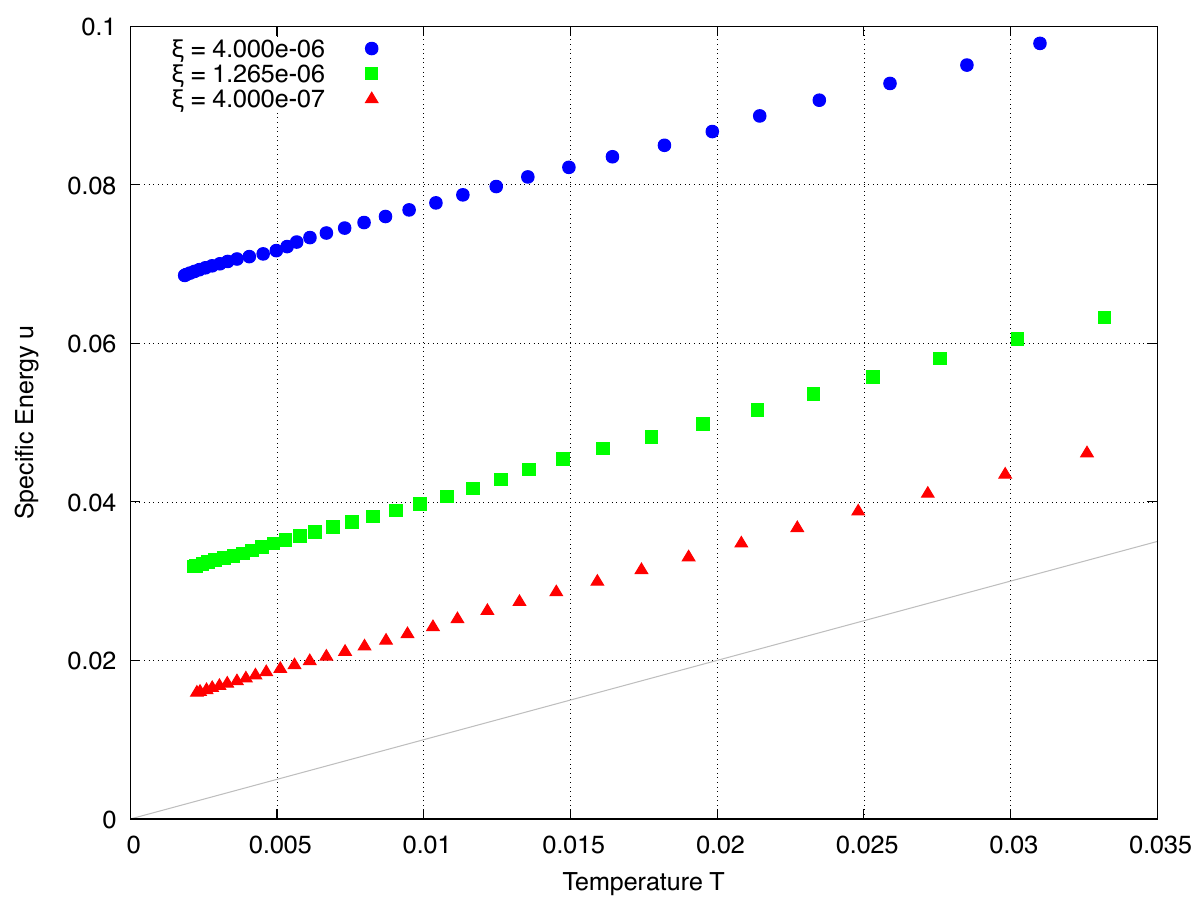}
    \caption{\label{fig:5} Cooling process for different value of the
      cooling rate, for temperature below $0.035$. Here the particles
      number is $N+2 = 200$.  }
\end{figure}

We would follow such a prescription, but a problem presents, i.e.,
during the cooling process, when the velocities of the gas particles
are diminished, the trajectories we obtain are not a good sample of
the Maxwell distribution, or better, a good sample of the Maxwell
distribution for the target temperature $T_{new}$. This can be
checked, for example, by looking at Figure \ref{cooling_process},
where in the upper panel it is reported, for a single trajectory, the
time average $\big\langle H_{FPU}\big\rangle/N$ of the FPU specific
energy $u$ versus temperature $T$, i.e., the double of the time
averaged gas kinetic energy per particle $\big\langle
K_{gas}\big\rangle/N$. We see that there is a trend versus a
decreasing of the averaged gas kinetic energy, but the process is in
no way monotonous, i.e., sometimes the kinetic energy increases
instead of diminishing. This is due to the fluctuations of the kinetic
energy with respect to the average, fluctuations which are still
large, in view of the fact that we cannot afford to reach values of
$N$ sufficiently large (we use up to $N+2=1000$
particles). Figure~\ref{cooling_process} refers to a case with
$N+2=200$, but this occurs for all the cases that we have checked.

To overcome this problem, the averages are performed in this way: we
choose a number $M=128$ of orbits and perform $M$ cooling processes as
described in the former Section. We compute the time averages during
the equilibration stage, first allowing the system to equilibrate for
a time interval of length $\frac{3}{5}\Delta t$, and then computing
the time average on a time interval of length $\frac{2}{5}\Delta t$.

So we end up with a number of points $(T_k^j,u_k^j)$ in the plane
$(T,u)$, where $j=1,\ldots,M$ refers to the $j$--th orbit, while
$k=1,\ldots,k_{max}$ refers to the $k$--th step in the cooling
process. Such points seem to cluster in the plane $(T,u)$, as shown in
the lower panel of Figure~\ref{cooling_process} where the black points
are determined by performing a moving average. This moving average is
computed after sorting all data points by temperature and convolving
the resulting sequence of $u$--values with a uniform kernel of fixed
width $w$. In practice, this amounts to replacing each point with the
arithmetic mean of its $w$ nearest neighbors along the $T$ axis. Such
black points appear to lie on a curve that continuously varies with
$T$.

We also implement a different averaging procedure: we divide the
temperature axis $T$ into bins whose widths are chosen such that each
bin contains $128$ points $(T_k^j,u_k^j)$, and then we average both
the temperature and the specific energy within each bin. Unlike the
bin--averaging procedure, the moving average does not introduce a
coarse graining of $T$ into discrete bins and preserves the original
temperature sampling. As a result, it provides a more continuous
representation of the underlying trend, at the cost of a residual
correlation between neighboring averaged points. Conversely, the
bin--averaging procedure produces a set of statistically independent
points, which is particularly useful for quantitative analyses such as
error estimation or fitting, where uncorrelated data are
preferable. The results of the two methods are consistent with each
other. The averaged data, as reported in Figure~\ref{cooling_process},
show that for temperatures above $T \approx 0.4$, the non linear
contributions are noticeable, while below we have $u\approx T$ as
predicted by the canonical ensemble for a system of oscillators.

As we remarked in the introduction, the FPU system begins to show an
ordered behavior for specific energy below the threshold $V_0/27$,
i.e., let us say for temperature below $T = 0.035$, which cannot be
appreciated on the scale used in Figure~\ref{cooling_process}. So,
from now on, we concentrate on temperatures below $T = 0.035$ and the
x-axis scale of the figures will be changed accordingly.

The first results are illustrated in Figure~\ref{fig:4}, where the
curve $u$ is reported as a function of $T$ for three different values
of the size of the FPU system, i.e. for $N+2=100$, $178$ and $316$. As
can be seen, in all cases the energy $u$ is well above the equilibrium
curve $u=T$, which means that the FPU system deviates from
equilibrium.

\begin{figure}[htbp]
    \includegraphics[width=\columnwidth]{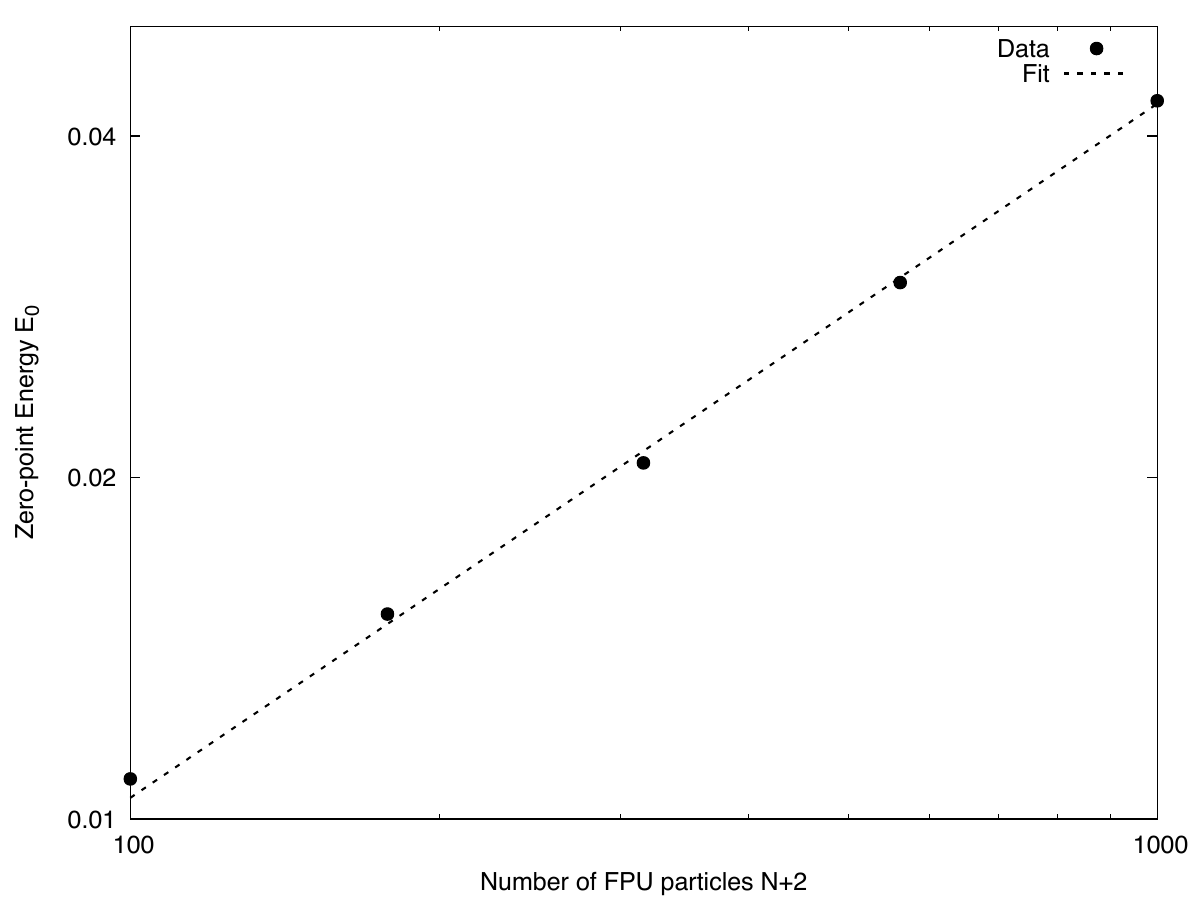}
    \caption{\label{fig:6} Dependence of the zero--point energy $E_0$
      from the particles number $N$. Points are computed by the
      bin-average procedure.}
\end{figure}

\begin{figure}[htbp]
    \includegraphics[width=\columnwidth]{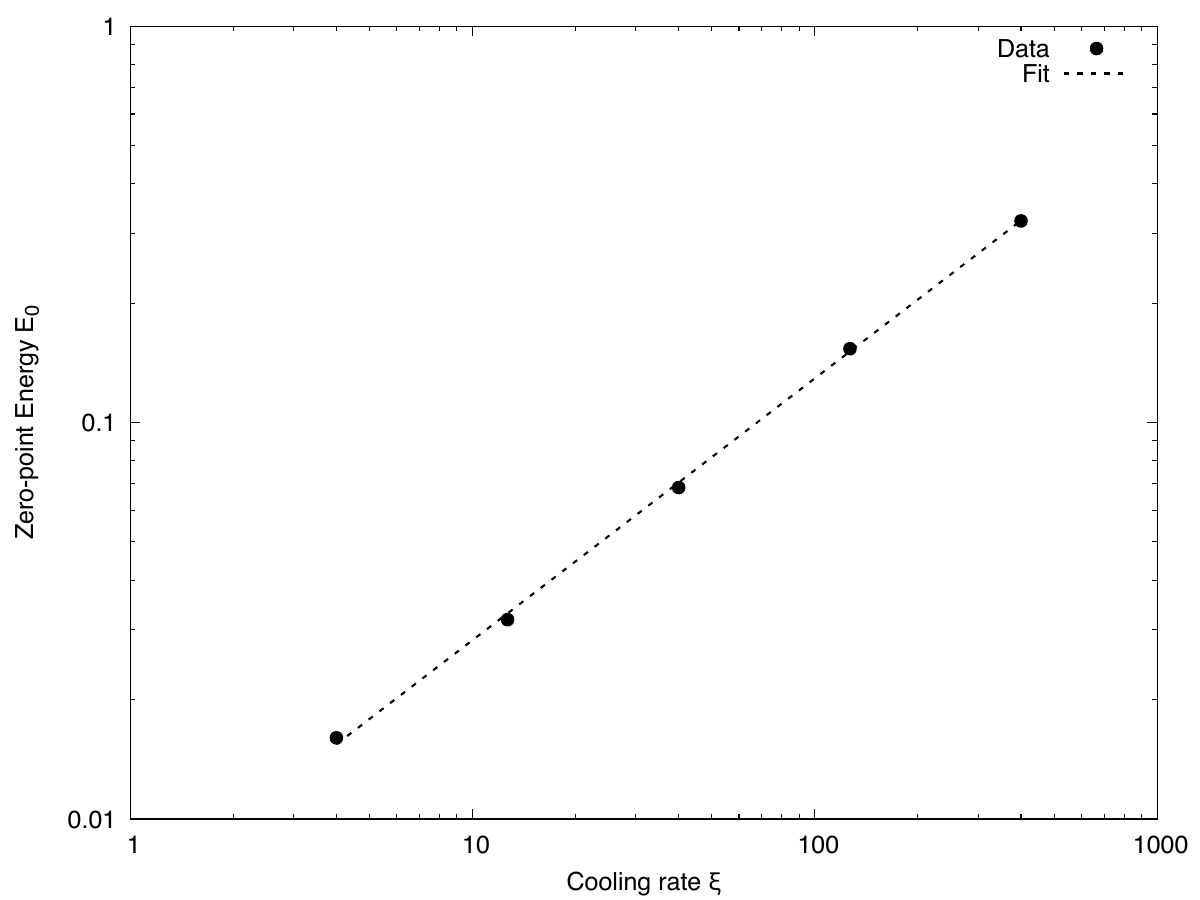}
    \caption{\label{fig:7} Dependence of the zero--point energy $E_0$
      from the cooling rate $\xi$. Cooling rate is rescaled by a
      factor $10^{-7}$. Points are computed by the moving average
      method.  }
\end{figure}

If one looks at each individual orbit and not at the average over the
sample, one can see that the point $(T_k^j,u_k^j)$ (at fixed $j$)
seems to lie on a curve having a slope vanishing at the lowest
temperatures. This would imply that the heat capacity is reduced by
reducing the temperature, which should explain why the energy $u$ lies
above the curve $u=T$. Unfortunately, our methods of computing the
averages blur such a phenomenon.

In any case, it seems that there is a part of the FPU energy that is
not exchanged with the gas. This energy $E_0$, which will be called
the zero--point energy, can be defined, by a first approximation, as
the value of $u$ at the lowest temperature reached. The Figure shows
the interesting phenomenon that the zero--point energy
\emph{increases} as the particles number $N$ increases.

The energy $E_0$ obviously depends on the cooling rate: in fact, if
one expects an \enquote{infinite time} before lowering the gas
temperature, the FPU system energy would lie on the curve $u=T$ (apart
from corrections due to the non linearity of the potential, negligible
at the lower temperatures). So, for a vanishing cooling rate, the
zero--point energy $E_0$ also vanishes. For small, but finite, cooling
rates, $E_0$ would no longer vanish, but it is reasonable to assume
that it would keep \emph{diminishing} as the cooling rate
decreases. So we perform different cooling processes with increasing
equilibration time $\Delta t$. In Figure~\ref{fig:5}, the curves $u$
vs. $T$ are reported for different cooling rates for a system of
$N+2=200$ particles. One can check that zero--point energy decrease
with the increasing of the cooling rate $\xi$ (defined as $\Delta T /
\Delta t$).

We could estimate the dependence of $E_0$ both from the particle
number $N$ and from the cooling rate $\xi$ by plotting in log--log
scale $E_0$ vs. $N$ (see Figure~\ref{fig:6}) and $E_0$ vs. $\xi$ (see
Figure~\ref{fig:7}). The points seem to lie on a straight line in both
cases, so it appears that the dependence follows a power law. The
slopes are quite similar in both cases, and in fact, we find $E_0\sim
N^{0.61}$ for a fixed $\xi = 4 \cdot 10^{-7}$ and $E_0\sim \xi^{0.66}$
for a fixed $N+2=200$ (both exponents have an accuracy of $0.01$). If
for simplicity we consider equal the two exponents, our computations
seem to suggest the possible law

\begin{equation}\label{eq:5}
    E_0\sim \big ( \xi N \big)^{2/3} \ .
\end{equation}

\section{\label{sec4}Conclusions}
So, the numerical calculations show that even in a model as simple as
the FPU one, it is not easy to understand the details of the cooling
process. In fact, if it is obvious that in a real process not all
available energy can be exchanged, the dependence of the zero--point
energy $E_0$ on the number of particles $N$ with an exponent close to
$2/3$ is unexpected and at the moment not understood. In the same way,
the dependence of $E_0$ on the cooling rate $\xi$ is not understood,
as it is not understood why it appears to be nearly identical to the
dependence on $N$.

In any case, the relation (\ref{eq:5}) shows that the cooling rate
must vanish faster than $1/N$ in order for the zero--point energy
$E_0$ to vanish in the thermodynamic limit. So, if this phenomenon
persists also in realistic models of the interaction between solids
and gases, it could be relevant for actual experiments. In fact, the
cooling rate could be so small for a macroscopic body that any real
cooling process would result in a non--negligible zero--point energy
thus affecting the thermodynamic behavior of the body.

\bibliography{bibliography}

\end{document}